\documentclass[floats,floatfix,showpacs,amssymb,prd,nofootinbib,superscriptaddress,aps,notitlepage,reprint]{revtex4-1}
\pdfoutput=1
\usepackage{graphicx}
\usepackage[utf8,latin1]{inputenc}
\usepackage[english]{babel}
\usepackage{graphicx, epsfig, amssymb} 
\usepackage{amsmath, amsfonts}
\usepackage{bm} 
\usepackage[normalem]{ulem} 
\usepackage{soul}
\usepackage[linktocpage]{hyperref}
\usepackage[caption=false]{subfig}
\usepackage[usenames]{color}

\def\be{\begin{equation}}
\def\ee{\end{equation}}

\newcommand{\diff}{\textrm{d}}

\newcommand{\bea}{\begin{eqnarray}}
\newcommand{\eea}{\end{eqnarray}}
\newcommand{\ben}{\begin{enumerate}}
\newcommand{\een}{\end{enumerate}}
\newcommand{\bi}{\begin{itemize}}
\newcommand{\ei}{\end{itemize}}

\def\ga{\mathrel{\raise.3ex\hbox{$>$\kern-.75em\lower1ex\hbox{$\sim$}}}}
	\def\la{\mathrel{\raise.3ex\hbox{$<$\kern-.75em\lower1ex\hbox{$\sim$}}}}

\def\be{\begin{equation}}
\def\ee{\end{equation}}

\def\I_M{{I_{\scriptscriptstyle M\times M}}}

\def\be{\begin{equation}}
\def\ee{\end{equation}}
\def\bea{\begin{eqnarray}}
\def\eea{\end{eqnarray}}
\newcommand{\beq}{\begin{eqnarray}}
\newcommand{\eeq}{\end{eqnarray}}

\newcommand{\beqal}{\begin{eqnarray}\label}
\newcommand{\none}{\end{eqnarray}}
\newcommand{\beqa}{\begin{eqnarray}}
\newcommand{\eeqa}{\end{eqnarray}}

\begin{document}
\title{\large The Gibbons-Hawking radiation of gravitons in the Poincar\'e and static patches of de~Sitter spacetime}

\author{Rafael P. Bernar}\email{rafael.bernar@icen.ufpa.br}
\affiliation{Faculdade de F\'{\i}sica, Universidade
Federal do Par\'a, 66075-110, Bel\'em, Par\'a, Brazil.}
\affiliation{Department of Mathematics, University of York, YO10 5DD, Heslington, York, United Kingdom.}

\author{Lu\'is C. B. Crispino}\email{crispino@ufpa.br}
\affiliation{Faculdade de F\'{\i}sica, Universidade
Federal do Par\'a, 66075-110, Bel\'em, Par\'a, Brazil.}

\author{Atsushi Higuchi}\email{atsushi.higuchi@york.ac.uk}
\affiliation{Department of Mathematics, University of York, YO10 5DD, Heslington, York, United Kingdom.}

\begin{abstract}	
We discuss the quantization of linearized gravity in the background de~Sitter spacetime using a gauge-invariant formalism to write the perturbed gravitational field in the static patch. This field is quantized after fixing the gauge completely. The response rate of this field to monochromatic multipole sources is then computed in the thermal equilibrium state with the well known Gibbons-Hawking temperature. We compare this response rate with the one obtained in the Bunch-Davies-like vacuum state defined in the Poincar\'e patch. These response rates are found to be the same as expected.  This agreement serves as a
verification of the infrared finite graviton two-point function in the static patch of de~Sitter spacetime found previously. 
\end{abstract}


\date{\today}

\maketitle


\section{Introduction}

Physics in de~Sitter spacetime is an interesting subject in its own right but it has increased its importance because 
the Universe's early stage of expansion is believed to have happened in a de~Sitter-like phase~\cite{Kazanas:1980tx,Sato:1980yn,Guth:1980zm,Linde:1981mu,Albrecht:1982wi}. Moreover, the accelerated 
expansion of our Universe~\cite{riess} means that de~Sitter spacetime is likely to approximate its late stages of 
evolution as well.


It is well known that the graviton two-point function is divergent in the infrared (IR) in the 
synchronous-transverse-traceless
gauge in the Poincar\'e patch, or the spatially-flat patch, of de~Sitter spacetime~\cite{Ford:1977dj}.  
These divergences arise
because the graviton mode functions reduce to those of the massless minimally-coupled scalar field that suffers from IR
divergences~\cite{Ford:1977in}.  In fact it is known that there is no Hadamard state invariant under the de~Sitter group
for massless minimally-coupled scalar field in de~Sitter spacetime~\cite{Allen:1985ux}.  It has been claimed that there
is no de~Sitter-invariant vacuum state for linearized gravity because of these and other
IR divergences [see, e.g.\ Refs.~\cite{Antoniadis:1985pj,Antoniadis:1986sb,Miao:2010vs,Kitamoto:2014gva,PhysRevD.96.106009}].
However, since the 
gravitational field is a gauge field unlike the scalar field, it is possible that these IR divergences can be a gauge artifact.

Indeed it has been shown that the IR-divergent part of the graviton two-point function mentioned above can be expressed
in a pure-gauge form~\cite{Allen:1986dd,Higuchi:1986py,Higuchi:2000ye}.  
More recently, it was shown that the graviton mode functions
can be modified by large gauge transformations corresponding to global shear transformations to make the two-point
function IR finite and, hence, de~Sitter invariant~\cite{Higuchi:2011vw}.  Some authors object by asserting that a
large gauge transformation, which by definition affects spatial infinity, would change 
physics~\cite{Miao:2011ng,Woodard:2015kqa}. However, as pointed out in Ref~\cite{Higuchi:2011vw}, a large gauge
transformation is equivalent to a local one as long as one is interested only in local physics.

It is also interesting to point out that the graviton two-point function constructed in the
hyperbolic patch~\cite{Hawking:2000ee}, global patch~\cite{higuchiweeks} and static patch~\cite{hbc1,ref:hbcproc} 
are all
IR finite.  These IR-finite two-point functions are consistent with the fact that the IR divergences in the two-point function
constructed in the Poincar\'e patch can be gauged away by (large) gauge transformations.

Now, the Bunch-Davies, or Euclidean, vacuum state~\cite{Chernikov:1968zm,Schomblond:1976xc,Bunch:1978yq} is
a thermal state of temperature $H/2\pi$, where $H$ is the Hubble constant for the de~Sitter 
expansion, with respect to the energy corresponding to the time translation in the static patch~\cite{gibbonshawking}.  
This fact, which we call the Gibbons-Hawking effect, is closely related to 
the Hawking radiation~\cite{ref:hawkingradiation} and the Unruh effect~\cite{unruhvacuum,crispinormp}.
Strictly speaking, the Gibbons-Hawking effect has not been shown for the
graviton field, but the two-point function of Refs.~\cite{hbc1,ref:hbcproc} was found assuming this effect.  
That is, this two-point function is for the thermal state of gravitons with temperature $H/2\pi$ 
in the static patch of de~Sitter spacetime.

In this paper we verify that the Bunch-Davies-like state for the graviton field 
in the Poincar\'e patch of de~Sitter spacetime, which has an
IR-divergent two-point function, is indeed the thermal equilibrium state with temperature $H/2\pi$ in the static patch,
which has an IR-finite two-point function.  We do so by  showing that a conserved multipole point source responds to 
the graviton field in the Bunch-Davies-like state as if it was placed in a thermal bath of temperature $H/2\pi$ with respect to
the energy corresponding to the time translation in the static patch. Similar calculations have been done for the scalar
and vector fields in Ref.~\cite{higuchi}. Similar comparisons between response rates of sources in
Schwarzschild spacetime have
also been made in the context of the Hawking and Unruh effects
in Refs.~\cite{Higuchi:1996aj,Higuchi:1998qc,Crispino:1998hp,Crispino:2000jx}.

The rest of the paper is organized as follows. 
In Sec.~\ref{sec:gravitondesitter} we describe the linearized gravitational field (gravitational perturbations) in 
$(3+1)$-dimensional de~Sitter spacetime and present the mode functions for these perturbations in spherical polar
coordinates in the Poincar\'e patch.
In Sec.~\ref{sec:quantization}, we describe our method of quantization of the gravitation field and determine the 
normalization constants for the modes found in Sec.~\ref{sec:gravitondesitter} such that the annihilation and creation
operators satisfy the standard commutation relations.  We also review  the quantization of the linearized gravitational field in the static patch
presented in Refs.~\cite{hbc1,ref:hbcproc}.  In Sec.~\ref{sec:gibbons-hawking} we verify the Gibbons-Hawking effect
for the gravitational field by comparing the response rates to a conserved multipole source in the Bunch-Davies-like state in
the Poincar\'e patch and in the thermal equilibrium with temperature $H/2\pi$ in the static patch. We conclude 
this paper with some remarks in Sec. \ref{sec:finalremarks}. In Appendix~\ref{appendix:expansion} we
present a derivation of the
expansion of the gravitational plane wave in terms of the modes in spherical polar coordinates.
Throughout this paper we use the metric signature $-+++$ and natural units such that $G=c=\hbar=k_B=1$.

\section{Gravitational perturbations in the Poincar\'e patch of de~Sitter spacetime}
\label{sec:gravitondesitter}

\subsection{Background de~Sitter Spacetime} 
\label{subsec:backgrounddesitter}

The line element covering the expanding half of de~Sitter spacetime (Poincar\'e patch) is given by:
\beqa
\diff s^2 = -\diff \tau^2 + e^{2 H \tau} \left(\diff \rho^2 + \rho^2 \diff \Omega_2^2\right),
\eeqa
where 
\beq
\diff \Omega_2^2=\gamma_{ij} \diff \hat{x}^i \diff \hat{x}^j= \diff \theta^2 + \sin^2 \theta \diff \phi^2
\eeq
is the line element on the unit 2-sphere. We reserve the letters from the Latin alphabet starting from $i,j,k,...$ to denote 
angular components. The (metric compatible) covariant derivative on the 2-sphere is denoted by $\hat{D}_i$. We also 
indicate any quantity on the 2-sphere with a hat over it.  
The line element which describes the static patch of de~Sitter spacetime reads
\beqa
\diff s^2 = -\left(1-H^2 r^2 \right) \diff t^2+\frac{\diff r^2}{1- H^2 r^2}+ r^2\diff \Omega_2^2, 
\eeqa
where the coordinates $t$ and $r$ are given in terms of the coordinates $\tau$ and $\rho$ as follows:
\beqa
r&=&\rho e^{H\tau}, \\
t&=&\tau-\frac{1}{2H}\ln \left(1-\rho^2e^{2H\tau}\right).
\eeqa
The Hubble constant $H$ is related to the cosmological constant $\Lambda$ by $\Lambda=3H^2$.

\subsection{Linearized Gravity in the Poincar\'e patch of de~Sitter spacetime} 
\label{subsec:linearizedgravity}

The Einstein-Hilbert action with a cosmological constant term is given by

\beq
S_{EH}=\frac{1}{16 \pi G}\int \sqrt{-\tilde{g}}(\tilde{R}-2\Lambda)d^4 x.
\eeq
The action for gravitational perturbations in a background spacetime (or linearized gravity) can be obtained by expanding the action $S_{EH}$ about a background metric, i.e. by writing 
$\tilde{g}_{\mu\nu}=g_{\mu\nu}+\sqrt{32\pi G}\,h_{\mu\nu}$, 
and retaining only terms of second order in $h_{\mu\nu}$\footnote{The zeroth order term is the Einstein-Hilbert action for the background solution and the linear term is a total derivative.}. In our case, the background metric $g_{\mu\nu}$ is the de~Sitter metric and we obtain the following quadratic Lagrangian:
\beqa
\mathcal{L}&=&\sqrt{-g}\left[\nabla _{\mu} h^{\mu \lambda}\nabla^{\nu}h_{\nu\lambda}-\frac{1}{2}\nabla_{\lambda}h_{\mu\nu}\nabla^{\lambda}h^{\mu\nu} \right. \nonumber \\
&&+\left. \frac{1}{2}(\nabla^{\mu} h - 2\nabla_{\nu}h^{\mu\nu})\nabla_{\mu} h -H^2\left(h_{\mu\nu}h^{\mu\nu}+\frac{h^2}{2}\right)\right], \nonumber \\ \label{eq:lagrangiandensity}
\eeqa
where $h=g^{\mu\nu}h_{\mu\nu}$.
The resulting Euler-Lagrange field equation is
\beqa
&& h_{\mu\nu}-2\nabla_{(\mu |}\nabla_{\lambda} {h^{\lambda}}_{|\nu)}+ g_{\mu\nu}\nabla_{\lambda}
\nabla_{\sigma}h^{\lambda \sigma} \nonumber \\
&&+\nabla_{\mu}\nabla_{\nu}h - g_{\mu\nu}\Box h-2H^2\left(h_{\mu\nu}+\frac{1}{2}g_{\mu\nu}h\right)=0. 
 \label{eq:fieldequations}
\eeqa

Due to the general coordinate invariance of the full Einstein-Hilbert action, the linearized theory is invariant under the gauge transformation 
\beq
h_{\mu\nu} \to h'_{\mu\nu}=h_{\mu\nu}+\nabla_\mu \xi_\nu + \nabla_\nu \xi_\mu,
\eeq
provided that the background spacetime is a vacuum solution to the Einstein's field equation with a cosmological constant~\cite{ref:Stewart49}.
We can choose a gauge such that $h=\nabla_{\mu}h^{\mu\nu}=0$, which greatly simplifies the equation of motion to
\beq
\left(\Box -2H^2\right)h_{\mu\nu}=0 \label{eq:gaugefixfieldeqs}.
\eeq
(See, e.g.\ Ref.~\cite{Higuchi:1991tn} for justification of this gauge.)
We shall find the solutions to Eq.~(\ref{eq:gaugefixfieldeqs}) in the Poincar\'e patch in spherical polar coordinates.  Thus,
we expand the field $h_{\mu\nu}$ in terms of harmonic tensors, following Refs.~\cite{kis2000,kodamaishibashi}. In $3+1$ dimensions, there will be metric perturbations (i) of the scalar type, for which the angular dependence comes from the scalar spherical harmonics and their covariant derivatives, and (ii) of the vector type, with angular dependence described by vector spherical harmonics and their covariant derivatives. Additionally, there are perturbations of the so-called tensor type, with angular dependence described by rank $2$ tensor spherical harmonics in higher dimensions. 
However, as is well known, there are no rank $2$ tensor spherical harmonics on the $2$-sphere~\cite{Higuchi:1986wu}, and hence we do not need to consider them here. The scalar spherical harmonics and their derivatives are orthogonal to the vector spherical harmonics and their derivatives with respect to the integration on the unit $2$-sphere. 

In the Poincar\'e patch, it is convenient to make an additional gauge choice in which $h_{\mu\tau}=0$. This set of gauge conditions is called the synchronous-transverse-traceless (STT) gauge.  (This gauge choice is possible because
the $h_{\mu\tau}$ component comes from transverse-traceless solutions to Eq.~(\ref{eq:gaugefixfieldeqs}) of the
pure-gauge form $h_{\mu\nu} = \nabla_\mu \xi_\nu + \nabla_\nu\xi_\mu$.) 

It follows that the non-vanishing positive-frequency\footnote{The meaning of ``positive-frequency'' will be clarified later.} components of the scalar-type perturbations, satisfying the gauge constraints, read
\beqa
h^{(S;klm)}_{\rho \rho}&=&\frac{A_S^{kl}}{\rho^2}\Phi^{kl}(\tau,\rho)\mathbb{S}^{(lm)}, \label{eq:scalarpertrhorho} \\
h^{(S;klm)}_{\rho i}&=&-\frac{A_S^{kl}\mathbb{S}^{(lm)}_i}{k_S}\left(\frac{\partial}{\partial \rho}+\frac{1}{\rho}\right)\Phi^{kl}(\tau,\rho), \label{eq:scalarpertrhoi} \\
h^{(S;klm)}_{ij}&=&A_S^{kl}\mathbb{S}^{(lm)}_{ij}\Psi^{kl}(\tau,\rho)-\frac{A_S^{kl}}{2}\gamma_{ij}\Phi^{kl}(\tau,\rho)\mathbb{S}^{(lm)}, \nonumber \\ \label{eq:scalarpertij}
\eeqa
where $\mathbb{S}^{(lm)}=\mathbb{S}^{(lm)}(\theta,\phi)$ are the scalar spherical harmonics, which satisfy
\beq
[\hat{D}_i\hat{D}^i+k_S^2]\mathbb{S}^{(lm)}(\theta,\phi)=0. \label{eq:scalarharmonics}
\eeq
The eigenvalues $k_S^2$ are
\beq
k_S^2=l(l+1), \ \ \ l=0,1,2,\ldots
\eeq
Solutions to Eq.~(\ref{eq:scalarharmonics}) are given by
\beqa \label{Slm-def}
\mathbb{S}^{(lm)}(\theta,\phi)=
\sqrt{\frac{(2l+1)}{4 \pi}\frac{(l-|m|)!}{(l+|m|)!}}P_{l}^{|m|}(\cos{\theta})e^{i m \phi}. \nonumber\\
\eeqa
The tensors $\mathbb{S}^{(lm)}_i(\theta,\phi)$ and $\mathbb{S}^{(lm)}_{ij}(\theta,\phi)$ are given by 
\beq
\mathbb{S}^{(lm)}_i(\theta,\phi)=-\frac{\hat{D}_i\mathbb{S}^{(lm)}(\theta,\phi)}{k_S}
\eeq
and
\beq
\mathbb{S}^{(lm)}_{ij}(\theta,\phi)=\left[\frac{\hat{D}_i \hat{D}_j}{k_S^2}+\frac{1}{2}\gamma_{ij}\right]\mathbb{S}^{(lm)}(\theta,\phi).
\eeq
The field $\Phi^{kl}(\tau,\rho)$ is a master variable and $\Psi^{kl}(\tau,\rho)$ reads
\beqa
\Psi^{kl}=\frac{2\rho^2}{(l-1)(l+2)}\left[\frac{\partial^2}{\partial \rho^2}+\frac{3}{\rho}\frac{\partial}{\partial \rho}-\frac{(l-1)(l+2)}{2\rho^2}\right]\Phi^{kl}. \nonumber \\
\eeqa
The normalization constants $A_S^{kl}$ will be determined later.

It is not possible to find $h_{\mu\nu}$ satisfying the STT gauge conditions in this form if $l=0$ or $1$.  
There are solutions
with $l=0,1$ which are not in this form, but they are either singular at the origin or of pure-gauge form.  
Thus, we only need to
consider the values of $l$ larger than or equal to $2$.  To emphasize this point we have outlined, in Appendix \ref{appendix:expansion}, the expansion of the
gravitational plane wave in terms of the modes in spherical polar coordinates, where only the modes with $l \geq 2$
are present.

The non-vanishing components of the vector-type metric perturbations can be written as
\beqa
h^{(V;k l m)}_{\rho i}&=&A^{kl}_V \Phi^{kl}(\tau,\rho)\mathbb{V}^{(l m)}_i,  \label{eq:vectorpertrhoi} \\
h^{(V;k l m)}_{ij}&=&-\frac{2k_VA^{kl}_V \rho^2\mathbb{V}^{(lm)}_{ij}}{(l-1)(l+2)}\left(\frac{\partial}{\partial \rho}+\frac{2}{\rho}\right) \Phi^{kl}(\tau,\rho). \nonumber \\ \label{eq:vectorpertij}
\eeqa
The vector spherical harmonics satisfy
\beq
(\hat{D}_j\hat{D}^j+k_V^2)\mathbb{V}^{(lm)}_i=0, \ \ \ \hat{D}^{i}\mathbb{V}^{(lm)}_i=0, \label{eq:vectorharmonics}
\eeq
with
\beq
k_V^2=l(l+1)-1,\ \ \ l=1,2,3,\ldots.
\eeq
The tensor $\mathbb{V}_{ij}^{(lm)}$ is written as
\beq
\mathbb{V}_{ij}^{(lm)}=-\frac{1}{2k_V}(\hat{D}_i\mathbb{V}_j+\hat{D}_j\mathbb{V}_i).
\eeq
On the unit $2$-sphere, one can write solutions to Eqs. (\ref{eq:vectorharmonics}) as
\beq
\mathbb{V}_{i}^{(lm)}(\theta,\phi)=\frac{\epsilon_{ij}}{\sqrt{l(l+1)}}\hat{D}^{j}\mathbb{S}^{(lm)}(\theta,\phi),
\label{Vsh-def}
\eeq
where $\epsilon_{ij}$ is the Levi-Civita tensor on $S^2$, defined by
\beqa
\epsilon_{\theta\theta}&=&\epsilon_{\phi\phi}=0, \\
\epsilon_{\theta\phi}&=&-\epsilon_{\phi\theta}=\sin{\theta}.
\eeqa
As in the scalar-type case there are no solutions to the gauge conditions of this form if $l=1$.  (There are no vector
spherical harmonics for $l=0$ as can be seen from the definition~\ref{Vsh-def}.)  For the same reason as for the 
scalar-type case, we only need to consider the case with $l\geq 2$.
The normalization constants $A_V^{kl}$ will be chosen later.

For the scalar- and vector-type perturbations to solve the equations of motion given by Eq.~(\ref{eq:gaugefixfieldeqs}), 
the master variable $\Phi^{kl}(\tau,\rho)$ takes the following form (or its complex conjugate or a linear combination of the 
two):
\beq
\Phi^{kl}(\tau,\rho)=\frac{k e^{\frac{H \tau}{2}}}{\sqrt{2H}}\mathrm{H}^{(1)}_{\frac{3}{2}}\left(\frac{k}{H}
e^{-H \tau}\right)j_l(k \rho), \label{eq:phieq}
\eeq
where $\mathrm{H}^{(1)}_{\frac{3}{2}}\left(\frac{k}{H}e^{-H\tau}\right)$ is the Hankel's function of the first kind, 
$j_{l}(k\rho)$ is the spherical Bessel function of the first kind, $k$ is a positive constant, and the overall constant factor 
has been chosen for later convenience.  The time-dependence of $\Phi^{kl}(\tau,\rho)$ is the same as that
for the plane-wave modes.

We now give a criterion to specify positive-frequency solutions in this setting. We require that $\Phi^{kl}$ for 
the positive-frequency solutions of the gravitational  perturbations to satisfy
\beq
\frac{\partial}{\partial \tau}\Phi^{kl} \approx -i ke^{-H\tau}\Phi^{kl},
\eeq 
in the limit $k \to \infty$. In other words, it should approach the positive-frequency solution in flat spacetime 
in the short wavelength limit\footnote{Note that the proper wave number is 
given by $ke^{-H\tau}$ in this case.}. Note that 
$\Phi^{kl}$ given in Eq.~(\ref{eq:phieq}) satisfies this requirement.  Now, one of the de~Sitter boosts,
$\tau\to \tau + \alpha$, $\rho \to e^{-\alpha}\rho$,  transforms the solution $\Phi^{kl}$ to $\Phi^{ke^{-\alpha},l}$.
Thus, once we choose the solutions~\eqref{eq:phieq} as the positive-frequency solutions for large 
$k$, we need to choose them as such for arbitrary $k$ to preserve the de~Sitter invariance of the set of positive-frequency
solutions, which leads to the de~Sitter invariance of the vacuum state 
(see, e.g.\ \cite{Higuchi:1986py}).  This choice of positive-frequency
solutions corresponds to the Bunch-Davies-like state, which is the standard choice of the vacuum~\cite{birrelldavies}.
From now on, we also set the Hubble constant to unity, i.e.\ $H=1$.

\section{Quantization of metric perturbations}
\label{sec:quantization}

To quantize the field $h_{\mu\nu}$, we follow a standard procedure outlined, for example, 
in Refs.~\cite{hbc1,hbc2}, which follow the general framework given in Ref.~\cite{Fewster:2012bj}.
We first define the symplectic product between two solutions of the equations of motion, given by Eq.~(\ref{eq:fieldequations}), to be
\beq
\Omega(h,h') \equiv \int_{\Sigma} \diff \Sigma n_{\alpha}(h_{\mu\nu}p'^{\alpha\mu\nu}-p^{\alpha\mu\nu}
h'_{\mu\nu}),
\eeq 
where $\Sigma$ is a Cauchy surface of a given patch of the spacetime 
with future-directed unit normal $n^{\alpha}$ and $p^{\alpha\mu\nu}$ is the 
conjugate momentum current defined by
\beq
p^{\alpha\mu\nu} \equiv \frac{1}{\sqrt{-g}}\frac{\partial \mathcal{L}}{\partial(\nabla_{\alpha}h_{\mu\nu})}.
\eeq
This symplectic product is independent of the choice of the Cauchy surface~\cite{Friedman:1978wla}. 

We choose the set of positive-frequency solutions given in Sec.~\ref{sec:gravitondesitter}, together with their 
complex conjugates, as a basis for the solutions to the free field equations (\ref{eq:fieldequations}) in the STT gauge. 
Then we define the inner product
\beq
\langle h,h' \rangle=- i\Omega(\overline{h},h'), \label{eq:innerproduct}
\eeq
where $\overline{h}_{\mu\nu}$ is the complex conjugate of $h_{\mu\nu}$. 
A positive- and a negative-frequency solutions are mutually orthogonal 
with respect to this inner product. Moreover, the inner product~(\ref{eq:innerproduct}) is positive definite on the space of 
positive-frequency solutions. Note that, since the STT gauge fixes the gauge completely, 
the symplectic product is non-degenerate. 
In other words, there are no solutions $h^{\mathrm{(null)}}_{\mu\nu}$ in the STT gauge satisfying 
$\Omega(h^{\mathrm{(null)}},h)=0$, for all solutions $h_{\mu\nu}$. (In our case it can readily be verified that
all such solutions to Eq.~(\ref{eq:fieldequations}) 
are pure-gauge solutions of the form $\nabla_\mu \xi_\nu + \nabla_\nu\xi_\mu$.)
Thus, we are considering only the space of physical solutions, i.e.\ all gauge degrees of freedom are eliminated, and the 
inner product~(\ref{eq:innerproduct}) is positive definite in the space of positive-frequency solutions.

\subsection{Quantization in the Poincar\'e patch}

The quantum field $h_{\mu\nu}$ (in the STT gauge) can be expanded as
\beqa
h_{\mu\nu}=\sum_{P,l,m}\int \diff k \left[a_{lm}^{P}(k)h_{\mu\nu}^{(P;klm)}+a_{lm}^{P}(k)^{\dagger}
\overline{h_{\mu\nu}^{(P;klm)}}\right], \nonumber\\ \label{eq:quantumexpansion}
\eeqa
where the label $P=S,V$ stands for scalar-type or vector-type perturbations, respectively, and the classical solutions 
$h_{\mu\nu}^{(P;klm)}$ are the positive-frequency solutions  given by Eqs.~(\ref{eq:scalarpertrhorho})-
(\ref{eq:scalarpertij}) and Eqs.~(\ref{eq:vectorpertrhoi}) and (\ref{eq:vectorpertij}). 
The canonical equal-time commutation relations are equivalent to
\beq
[a_{lm}^{P}(k),a_{l'm'}^{P'}(k')^{\dagger}]=\delta^{PP'}\delta^{ll'}\delta^{mm'}\delta(k-k')
\eeq
and
\beq
[a_{lm}^{P}(k),a_{l'm'}^{P'}(k')]=[a_{lm}^{P}(k)^{\dagger},a_{l'm'}^{P'}(k')^{\dagger}]=0,
\eeq
provided the complete set of positive-frequency solutions are normalized with respect to the inner 
product~(\ref{eq:innerproduct}), i.e.\ if
\beq
\langle h^{(P;klm)}, h^{(P';k'l'm')}\rangle = \delta^{PP'}\delta^{ll'}\delta^{mm'}\delta(k-k').
\eeq
Then, the vacuum $|0\rangle$, defined to be the state annihilated by all $a_{lm}^{P}(k)$, is the standard
Bunch-Davies-like state. 

In the STT gauge, the conjugate momentum current is simply
$p^{\alpha\mu\nu}=-\nabla^{\alpha}h^{\mu\nu}$,
so that the inner product can be written as
\beq
\langle h, h' \rangle = -i \int_{\Sigma} \diff \Sigma \left(h'_{\mu\nu}\partial_\tau \overline{h^{\mu\nu}}-\overline{h_{\mu\nu}}\partial_\tau h'^{\mu\nu}\right), \label{eq:poincareinnerproduct}
\eeq
where, in this case, $\Sigma$ is a $\tau=$ constant hypersurface. Using Eq.~(\ref{eq:poincareinnerproduct}) and the identities
\beq
\overline{\mathrm{H}^{(1)}_{\nu}(x)}\partial_x\mathrm{H}^{(1)}_{\nu}(x)-\mathrm{H}^{(1)}_{\nu}(x)\partial_x\overline{\mathrm{H}^{(1)}_{\nu}(x)}=\frac{4ie^{\pi \mathrm{Im} \nu}}{\pi x}
\eeq
and
\beq
\int_{0}^{\infty} \diff \rho\,\rho^2\, j_{l}(k\rho)j_{l}(k'\rho) = \frac{2}{\pi k^2}\delta(k-k'), \label{eq:besselproperty}
\eeq
one can readily compute the normalization constants for $h^{(P;klm)}_{\mu\nu}$, with $P=S$ and $V$, defined by 
Eqs.~\eqref{eq:scalarpertrhorho}-\eqref{eq:scalarpertij} and Eqs.~\eqref{eq:vectorpertrhoi} and \eqref{eq:vectorpertij},
respectively. After some cumbersome but straightforward computations, 
we obtain
\beq\label{AklV}
A^{kl}_V=\frac{1}{k}\sqrt{\frac{(l-1)(l+2)}{2}}
\eeq 
and
\beq\label{AklS}
A^{kl}_S=\frac{1}{k^2}\sqrt{\frac{(l-1)l(l+1)(l+2)}{2}}.
\eeq

\subsection{Quantization in the static patch}
In Refs.~\cite{hbc1,ref:hbcproc}, 
the quantization procedure outlined in the previous subsection was used to quantize the metric perturbations in the static 
patch of de~Sitter spacetime. We review it here for completeness. 

One can write the non-vanishing 
components of the (positive-frequency) scalar-type metric perturbations as
\beqa
h_{ab}^{(S;\omega l m)}&=&\mathbb{S}^{(l m)}\left( D_a D_b -\frac{1}{2}g_{ab}\Box \right)\left(r \psi_S^{\omega l}\right),
\nonumber \\
\label{eq:scalarhab-static}   \\
h_{ij}^{(S;\omega l m)}&=&\frac{r^{2}}{2}\gamma_{ij}\mathbb{S}^{(l m)}(\Box + 2)\left(r\psi_S^{\omega l}\right), \label{eq:scalarhij-static}
\eeqa
where $\psi_S^{\omega l}$ is the master field for this case (see Ref.~\cite{hbc1} for the details). The first letters of the Latin alphabet ($a,b,c,\ldots$) are used to denote components in the orbit spacetime spanned by the $t$ and $r$ coordinates, with metric
\beq
\diff s^2_{\mathrm{orbit}}=-\left(1- r^2 \right) \diff t^2+\frac{\diff r^2}{1-  r^2}. 
\eeq
The derivative operator $D_a$ is the covariant derivative on this spacetime.
The positive-frequency vector-type perturbations read
\beq
h_{ai}^{(V; \omega l m)}
=\epsilon_{ab}D^{b}\left(r\psi^{\omega l}_V\right)\mathbb{V}_i^{(l m)}, \label{eq:vectorhai-static}
\eeq
with all other components vanishing, where $\epsilon_{ab}$ is the Levi-Civita tensor in the orbit spacetime.

The master fields $\psi_S^{\omega l}$ and $\psi_V^{\omega l}$ are given by 
\beq
&&\psi_P^{\omega l}(t,r)=A_{\text{static}}^{P;\omega l}e^{-i\omega t} r^{l+1}(1-r^2)^{i\omega /2} \nonumber \\
&& \times F\left(\frac{1}{2}(i\omega +l+1),\frac{1}{2}(i\omega +l+2);l+\frac{3}{2};r^2\right), 
\label{eq:staticmasterfield}
\eeq
where $A_{\text{static}}^{P;\omega l}$ are normalization constants. Since we are in the static patch, the positive-frequency property is manifest with the factor $e^{-i\omega t}$.  One then expands the quantum field in the same manner as in Eq.~(\ref{eq:quantumexpansion}).  That is,
\beqa
h_{\mu\nu}=\sum_{P,l,m}\int \diff \omega \left[b_{lm}^{P}(\omega)h_{\mu\nu}^{(P;\omega lm)}+b_{lm}^{P}(\omega)^{\dagger}\overline{h_{\mu\nu}^{(P;\omega lm)}}\right]. \nonumber\\ \label{eq:quantumexpansionstatic}
\eeqa
By normalizing the classical fields $h_{\mu\nu}^{(P;\omega lm)}$ with respect to the inner product~(\ref{eq:innerproduct}), i.e.\ by letting
\beq
\langle h^{(P;\omega l m)},h^{(P';\omega'l'm')}\rangle = \delta^{PP'}\delta^{ll'}\delta^{mm'}\delta(\omega-\omega'),
\eeq
one obtains the usual  commutation relations between the operators $b_{lm}^{P}(\omega)$ and 
$b_{lm}^{P}(\omega)^{\dagger}$, i.e.
\beq
\left[ b_{lm}^P(\omega),b_{l'm'}^{P'}(\omega')^\dagger\right]
= \delta^{PP'}\delta^{ll'}\delta^{mm'}\delta(\omega-\omega'),
\eeq
with all other commutators vanishing.
The static vacuum $|0S \rangle$ is defined by requiring that it should be 
annihilated by all the annihilation operators $b_{lm}^{P}(\omega)$. By computing the inner 
product~(\ref{eq:innerproduct}) with the metric perturbations given in 
Eqs.~(\ref{eq:scalarhab-static})-(\ref{eq:scalarhij-static}) and Eq.~(\ref{eq:vectorhai-static}), the normalization constants 
are determined as follows~\cite{hbc1}:
\beqa
|A_{\text{static}}^{S;\omega l}|^2=\frac{\sinh \pi \omega \left|\Gamma\left(\frac{i\omega +l+2}{2}\right)
\Gamma\left(\frac{i\omega +l+1}{2}\right)\right|^2}{2 \pi^2 (l-1)l(l+1)(l+2)\left|\Gamma\left(l+\frac{3}{2}\right)
\right|^2} \nonumber \\
\eeqa
and
\beqa
|A_{\text{static}}^{V;\omega l}|^2=\frac{\sinh\pi\omega\left| \Gamma\left(\frac{i\omega+l+1}{2}\right)
\Gamma\left(\frac{i\omega+ l+ 2}{2}\right)\right|^2}{8\pi^2(l-1)(l+2)\left|\Gamma(l+\frac{3}{2})\right|^2}. 
\eeqa

\section{Response rate to a multipole external source}
\label{sec:gibbons-hawking}

\subsection{Response rate in the Poincar\'e patch}

Having obtained the normalized graviton modes, we introduce a multipole source term that couples to the field
$h_{\mu\nu}$ in the Lagrangian density (\ref{eq:lagrangiandensity}) as follows:
\beq
\frac{\mathcal{L}_{\mathrm{int}}}{\sqrt{-g}}=\frac{\sqrt{32 \pi}}{2}T^{\mu\nu}(x)h_{\mu\nu}(x), \label{eq:intlagrangian}
\eeq
where $T^{\mu\nu}$ is the energy-momentum tensor of the source. We note that, since $T^{\mu\nu}$ is a symmetric 
second rank tensor, one can expand it in the same way as the metric perturbations. Moreover, the coupling in the 
interaction term implies that products of scalar- and vector-type parts vanish when integrated on the whole spacetime.
Thus, we can consider separately each type of energy-momentum tensor which couples to the same type of graviton modes. 
Moreover, the energy-momentum tensor has to be conserved in the background spacetime, in order for the interaction Lagrangian given by Eq.~(\ref{eq:intlagrangian}) to be gauge invariant.
We construct the conserved scalar-type energy-momentum tensor 
$T^{\mu\nu}_{(S;Elm)}$ with the condition that $T^{\tau\mu}_{(S;Elm)} = 0$.  Then the conservation equation
$\nabla_\mu T^{\mu\nu}_{(S;Elm)}=0$ leads to the following nonzero components:
\beqa
T^{\rho\rho}_{(S;Elm)}&=&\frac{j^{El}_S}{\rho^2}\mathbb{S}^{(lm)}, \label{eq:scalarT_rhorho}\\
T^{i\rho}_{(S;Elm)}&=&-\frac{1}{k_S\rho^2}\left(\frac{\partial}{\partial \rho}+\frac{1}{\rho}\right)j^{El}_{S}\mathbb{S}^{i(lm)},\label{eq:scalarT_irho}\\
T^{ij}_{(S;Elm)}&=&g^{El}_{S}\mathbb{S}^{ij(lm)}-\frac{\gamma^{ij}j^{El}_{S}}{2\rho^{4}}\mathbb{S}^{(lm)},  \label{eq:scalarT_ij}
\eeqa
where
\beqa
g^{El}_{S}\equiv\left\{\frac{2\rho^{-2}}{(l+2)(l-1)}\left[\frac{\partial^2}{\partial \rho^2}+\frac{3}{\rho}\frac{\partial}{\partial \rho}\right]-\frac{1}{\rho^4}\right\}j^{El}_{S}. 
\eeqa
The function $j^{El}_{S}(\tau,\rho)$ is arbitrary and we will choose its form later.
 
The conserved vector-type energy-momentum tensor can be found under the same condition
$T^{\tau \mu}_{(V;Elm)} = 0$ as
\beqa
T_{(V;Elm)}^{\rho i}&=&\frac{j^{El}_{V}}{\rho^2}\mathbb{V}^{i(lm)}, \label{eq:vectorTrhoi} \\
T^{ij}_{(V;Elm)}&=&-\frac{2 k_V g^{El}_{V}}{\rho^{2}} \mathbb{V}^{ij(lm)}, \label{eq:vectorTij}
\eeqa
with all other components vanishing, where
\beq
g^{El}_V\equiv\frac{1}{(l+2)(l-1)}\left(\frac{\partial}{\partial \rho}+\frac{2}{\rho}\right)j^{El}_V.
\eeq
Note that this energy-momentum tensor satisfies the conservation condition because 
the $\mathbb{V}^{(lm)}_{ij}$ are traceless. The function $j_V^{El}(\tau,\rho)$ is arbitrary, as in the scalar-type case.

We now let 
\beqa
j^{El}_{P}(\tau,\rho)=\lim\limits_{\rho_0 \to 0}\lambda\frac{ e^{-(l+5+n_P)\tau}}{(l+2)!}
\left(-\frac{\partial}{\partial \rho}\right)^{l+2}\delta(\rho-\rho_0)e^{iE\tau}, \nonumber \\ \label{eq:jfunction}
\eeqa
where $n_S=2$, $n_V=1$, and $\lambda$ is a small coupling constant.  The number of $\rho$-derivatives has been 
chosen so that there is a nonzero but finite response rate for given angular momentum $l$.  The exponential factor 
$e^{-(l+5+n_P)\tau}$ has been chosen so that the response rate does not vary with $\tau$.

Let us now compute the response rate (probability of emission/absorption per unit time) of the graviton field in the vacuum 
to the multipole sources $T^{\mu\nu}_{(P:Elm)}$. 
If the initial state is the vacuum, there is only the possibility of emission, to lowest order in $\lambda$.
Due to the form of the sources given by Eqs.~(\ref{eq:scalarT_rhorho})-(\ref{eq:scalarT_ij}), in the scalar-type case, and by
Eqs.~(\ref{eq:vectorTrhoi})-(\ref{eq:vectorTij}), in the vector-type case, the only non-vanishing amplitudes 
 (to lowest order in perturbation theory) are the ones for the emission of a $P$-type graviton 
(when the initial state is the vacuum $|0\rangle$) with quantum numbers $k$, $l$ and $m$.  These amplitudes 
are given by
\beqa
\mathcal{A}^{P}_{klm}&=&i\langle 0|a^{P}_{lm}(k)\int 
\diff^4x \mathcal{L}_{\mathrm{int}}|0\rangle \nonumber \\
&=& i \int \diff \tau \int \diff \rho \diff \Omega_2 e^{3\tau} \rho^2 \overline{h^{(P;klm)}_{\mu\nu}} 
T_{(P;Elm)}^{\mu\nu}. \label{eq:emissionamp}
\eeqa
The response rate from the vacuum $| 0 \rangle$ is then~\cite{higuchi}
\beq
\mathcal{R}^{P;E}_{\text{Poincar\'e}}=\int_{0}^{\infty} \diff k \frac{|\mathcal{A}^{P}_{klm}|^2}{T_{\textrm{tot}}},
\eeq
where 
\beq
T_{\textrm{tot}} =2 \pi \delta(0)=\int_{-\infty}^{\infty} \diff \tau \label{eq:totaltime}
\eeq
is the total time as measured by the comoving observer (cf. Refs.~\cite{Crispino:1998hp,Higuchi:1998qc,PhysRevD.70.127504,crispino2,PhysRevD.84.025010} and references therein). 
The source is nonzero only at $\rho=0$. Therefore, we can 
use the following expansion around $\rho=0$ for the master field:
\beqa
\phi^{kl}(\tau,\rho) &\approx& \frac{\sqrt{\pi} k e^{\frac{\tau}{2}}\mathrm{H}^{(1)}_{\frac{3}{2}}(ke^{-\tau})}{2 ^{\frac{3}{2}}\Gamma\left(l+\frac{3}{2}\right)}\left[\left(\frac{k \rho}{2}\right)^l \right. \nonumber \\
&& \left. -\frac{\left(\frac{k \rho}{2}\right)^{l+2}}{\left(l+\frac{3}{2}\right)}+\frac{\left(\frac{k \rho}{2}\right)^{l+4}}{2\left(l+\frac{3}{2}\right)\left(l+\frac{5}{2}\right)}\right].
\eeqa
Using this expansion and Eq.~(\ref{eq:emissionamp}), 
we find that the squared transition amplitude, integrated over $k$, can be written as
\begin{widetext}
\beqa
\int \diff k|\mathcal{A}^{P}_{klm}|^2&=&  \frac{\pi \lambda^2 |k^{n_p} A_{P}^{kl}|^{-2} }{2^{2l+3} |\Gamma\left(l+\frac{3}{2}\right)|^2} \int \frac{\diff k}{k} \int \limits_{-\infty}^{\infty} \diff \tau 
  \int \limits_{-\infty}^{\infty} \diff \tau' (ke^{-\tau})^{l+n_P+\frac{3}{2}}(ke^{-\tau'})^{l+n_P+\frac{3}{2}} 
  H^{(1)}_{\frac{3}{2}}\left(ke^{-\tau}\right)\overline{H^{(1)}_{\frac{3}{2}}}\left(ke^{-\tau'}\right)e^{iE(\tau-\tau')}. \nonumber \\
\label{amplitude}
\eeqa
\end{widetext}
Note that the factor $|k^{n_P}A_{P}^{kl}|^{-2}$ does not depend on $k$ and, hence, it can be moved outside 
the integral.\footnote{The normalization factor squared, $|A_{P}^{kl}|^2$, appears in the numerator, but a factor 
proportional to $|A_{P}^{kl}|^4$ appears in the denominator.  
This explains the factor $|A_{P}^{kl}|^{-2}$ in Eq.~\eqref{amplitude}.}

Now, we make the following change of variables 
\beqa
T&=&\frac{\tau+\tau'}{2}, \\
\tau_r&=&\tau-\tau', \\
K&=&ke^{-\frac{\tau+\tau'}{2}},
\eeqa
so that the integrand does not depend on $T$ and the integral over this variable can be factored out. 
It will be cancelled by the total time [see Eq.~(\ref{eq:totaltime})] when we compute the response rate.  Thus, we find
\beqa
\mathcal{R}^{P;E}_{\text{Poincar\'e}}&=&\frac{\pi \lambda^2 |k^{n_p}A_{P}^{kl}|^{-2} }{2^{2l+3} |\Gamma\left(l+\frac{3}{2}\right)|^2} \int \frac{\diff K}{K} \int_{-\infty}^{\infty} \diff \tau_r K^{2l+2n_P+3} \nonumber \\ 
&& \times  H^{(1)}_{\frac{n+1}{2}}\left(Ke^{-\tau_r/2}\right)\overline{H^{(1)}_{\frac{n+1}{2}}}\left(Ke^{\tau_r/2}\right)e^{iE\tau_r}. \nonumber \\
\eeqa
We perform a further change of variables given by
\beqa
x&=&K e^{-\tau_r/2}, \\
y&=&K e^{\tau_r/2}.
\eeqa
We thus obtain
\beqa
\mathcal{R}^{P;E}_{\text{Poincar\'e}}&=&\frac{\pi \lambda^2 |k^{n_p} A_{P}^{kl}|^{-2} }{2^{2l+3} |\Gamma\left(l+\frac{3}{2}\right)|^2}\nonumber \\
&&\times \left| \int \limits_{0}^{\infty} \diff x x^{l+n_P+\frac{1}{2}+iE} H^{(1)}_{\frac{3}{2}}(x)  \right|^2.
\eeqa
Using Eq.~(A6) of Ref.~\cite{higuchi}, namely
\beqa
\int_0^{\infty+i\epsilon} z^\mu H_\nu^{(1)}(z)\mathrm{d} z
& = & \frac{2^\mu}{\pi}\exp\left[ \frac{1}{2}i(\mu - \nu)\pi\right] \nonumber \\
&& \times \Gamma\left( \frac{\mu+\nu+1}{2}\right)\Gamma\left( \frac{\mu - \nu + 1}{2}\right),\nonumber \\
\eeqa
for $\mathrm{Re}\,\mu - |\mathrm{Re}\,\nu| + 1 > 0$,
we find the following result:
\beqa
\mathcal{R}^{P;El}_{\text{Poincar\'e}}= \frac{ \lambda^2 e^{-\pi E} \left|\Gamma\left(\frac{l+iE+n_P+3}{2}\right)\Gamma\left(\frac{l+iE+n_P}{2}\right)\right|^2 }{4^{1-n_P} \pi |k^{n_p} A^{kl}_{P}|^2
\left|\Gamma\left(l+\frac{3}{2}\right)\right|^2}. \nonumber \\
\label{Poincare-rate}
\eeqa

\subsection{Response rate in the static patch}
We now compare the response rate in the Poincar\'e patch, Eq.~\eqref{Poincare-rate},
to the one obtained in the static patch from the same source in thermal equilibrium with temperature $1/2 \pi$, the 
Gibbons-Hawking temperature for de~Sitter spacetime (with $H=1$). 

We first assume $E > 0$.   Then
\beqa
\int \diff^4 x \sqrt{-g}\ T^{\mu\nu}_{(S;Elm)}h_{\mu\nu}& = &2\pi\lambda(l+1-iE)(l+3-iE) \nonumber \\
&& \times A_{\text{static}}^{S;\omega l}b^{S}_{l,-m}(E) 
\eeqa
and
\beqa
\int \diff^4 x \sqrt{-g}\ T^{\mu\nu}_{(V;Elm)}h_{\mu\nu} & = & 4\pi \lambda
(iE-l-2)\nonumber \\
&& \times A_{\text{static}}^{V;\omega l}b^{V}_{l,-m}(E).
\eeqa
If the initial state is given by a one-particle state $b^{P}_{l,-m}(\omega)^\dagger|0S \rangle$, $P=S$ or $V$,
in the static patch,
we find that the absorption probability per unit time is
\beqa
\mathcal{P}_{\mathrm{static}}^{S;\omega l,-m}&=&
	2 \pi \lambda^2 |l+1+iE|^2|l+3+iE|^2 \nonumber \\
&&\times	|A_{\text{static}}^{S;\omega l}|^2\delta(\omega-E), \label{eq:scalarabsprob-static}
\eeqa
in the scalar-type case, and
\beqa
\mathcal{P}_{\mathrm{static}}^{V;\omega l,-m}&=&
2 \pi \lambda^2 |l+2+iE|^2|A_{\text{static}}^{V;\omega l}|^2\delta(\omega-E), \label{eq:vectorabsprob-static}
\eeqa
in the vector-type case. Hence, in the scalar-type case the absorption rate in thermal equilibrium with temperature $1/2 \pi$ is
\beqa
\mathcal{R}^{S;El}_{\text{static}}&=& \int \mathcal{P}_{\mathrm{static}}^{S;\omega l,-m} \frac{\diff \omega}{e^{2\pi \omega}-1} \nonumber \\
&=&\frac{ 8\lambda^2 e^{-\pi E}  \left|\Gamma\left(\frac{l+iE+5}{2}\right)\Gamma\left(\frac{l+iE+2}{2}\right)\right|^2 }{ \pi (l-1)l(l+1)(l+2) \left|\Gamma\left(l+\frac{3}{2}\right)\right|^2}, \label{eq:scalarresponserate-static}
\eeqa
and the absorption rate in the vector-type case reads
\beqa
\mathcal{R}^{V;El}_{\text{static}}= \frac{ 2\lambda^2 e^{-\pi E} \left|\Gamma\left(\frac{l+iE+4}{2}\right)\Gamma\left(\frac{l+iE+1}{2}\right)\right|^2 }{ \pi (l-1)(l+2) \left|\Gamma\left(l+\frac{3}{2}\right)\right|^2}. \label{eq:vectorresponserate-static}
\eeqa
If $E < 0$, there is emission of a graviton by the source. The emission probabilities per unit time are again given by Eqs.~(\ref{eq:scalarabsprob-static}) and (\ref{eq:vectorabsprob-static}) with the change $E \to |E|$. However, in this case, we have to take into account both spontaneous and induced emissions. Hence, the emission rates are
\beq
\mathcal{R}^{P;El}_{\text{static}}&=& \int \mathcal{P}_{\mathrm{static}}^{P;\omega l,-m} \diff \omega\left(\frac{1}{e^{2\pi \omega}-1}+1\right).
\eeq
Thus, we find that the emission rates are again given by Eqs.~(\ref{eq:scalarresponserate-static}) and (\ref{eq:vectorresponserate-static}) (without the change $E\to |E|$).  By comparing these results with
Eq.~\eqref{Poincare-rate}, where $n_S=2$ and $n_V=1$, and where $A_P^{kl}$, with $P=S$ and $V$, are given by
Eqs.~\eqref{AklS} and \eqref{AklV}, respectively, we find 
$\mathcal{R}^{P;El}_{\text{static}} = \mathcal{R}^{P;El}_{\text{Poinacar\'e}}$ for both $P=S$ and $V$.

Thus, we have shown that the response rate of the vacuum $|0\rangle$ to the conserved 
external multipole sources $T^{\mu\nu}_{(P;Elm)}$, $P=S,V$, is identical to the response rate of the heat bath with 
temperature $1/2 \pi$ in the static patch.
\section{Concluding Remarks}
\label{sec:finalremarks}
In this paper we verified the Gibbons-Hawking effect, i.e.\ the fact that the standard 
vacuum state for quantum field theory
in de~Sitter spacetime is a thermal equilibrium state with temperature $H/2\pi$, where $H$ is the Hubble constant,
for the gravitational perturbations.  Although this was an expected result, it is reassuring to verify it explicitly.  
Strictly speaking, derivations of this and other related effects in general spacetimes with bifurcate Killing horizons~\cite{gibbonshawking,Kayreport,Sewell:1982zz} have been given only for non-gauge fields.
It would be interesting to close this gap and find a general derivation of this and other related effects applicable also to
gauge fields including perturbative gravity.  

Our result also serves as a check of the IR-finite graviton two-point function in the static patch found in 
Refs.~\cite{hbc1,ref:hbcproc}.  That is, we have verified explicitly that the standard vacuum state for the gravitational 
pertubations in the Poincar\'e patch, correponding to an IR-divergent two-point function, and the thermal state in the
static patch, corresponding to an IR-finite two-point function, have the same response to conserved external
energy-momentum sources. The conservation of the energy-momentum tensor also ensures gauge invariance of the response rates. This is an interesting first step for examining physics in de~Sitter spacetime using the
static patch, where the IR properties of the gravitational perturbations are better controlled. Since there have been
disagreement about the physical significance of the IR divergences in the Poincar\'e patch, it would be interesting to develop
gravitational perturbation theory in the static patch, now that the thermal state studied in Refs.~\cite{hbc1,ref:hbcproc} has
been shown to produce the correct physics when probed by an external source. 


\begin{acknowledgments}
We would like to acknowledge Conselho Nacional de Desenvolvimento Cient\'ifico e Tecnol\'ogico (CNPq) and Coordena\c{c}\~ao de Aperfei\c{c}oamento de Pessoal de N\'ivel Superior (CAPES). A. H. thanks the Universidade Federal do Par\'a (UFPA) in Bel\'em for the kind hospitality.

\end{acknowledgments}
\appendix

\section{Expansion of the gravitational plane wave in flat space in terms of the modes in spherical polar coordinates} 
\label{appendix:expansion}

In this Appendix we review the expansion of the gravitational plane wave in spatially-flat spacetime, including the 
Poincar\'e patch of de~Sitter spacetime, in terms of the modes in spherical polar coordinates.  This Appendix is included 
in order to emphasize that only the modes with $l \geq 2$ are present in the expansion of the gravitational plane waves
in the Poincar\'e patch.  We note that both  the plane-wave modes and the modes in spherical polar coordinates have 
the time-dependence given by $e^{\frac{H\tau}{2}}\mathrm{H}_{\frac{3}{2}}^{(1)}(ke^{-H\tau}/H)$ 
[see Eq.~\eqref{eq:phieq}].
Hence, it  is sufficient to consider the space-dependence of the plane waves and the vector- and scalar-type modes.
Thus,  we extract the space-dependent part of the scalar-type modes given by 
Eqs.~\eqref{eq:scalarpertrhorho}-\eqref{eq:scalarpertij} as
\beqa
H^{(S;klm)}_{\rho \rho}&=&\frac{A_S^{kl}}{\rho^2}j_l(k\rho)\mathbb{S}^{(lm)},\label{HSklm} \\
H^{(S;klm)}_{\rho i}&=&-\frac{A_S^{kl}\mathbb{S}^{(lm)}_i}{k_S}\left(\frac{\partial}{\partial \rho}+\frac{1}{\rho}\right)j_l(k\rho), \\
H^{(S;klm)}_{ij}&=&A_S^{kl}\mathbb{S}^{(lm)}_{ij}\psi^{kl}(\rho)-\frac{A_S^{kl}}{2}\gamma_{ij}j_l(k\rho)\mathbb{S}^{(lm)}, \nonumber \\
\eeqa
where
\beqa
\psi^{kl}(\rho) & = & 
\frac{2\rho^2}{(l-1)(l+2)}\nonumber \\
&& \times \left[\frac{\partial^2}{\partial \rho^2}+\frac{3}{\rho}\frac{\partial}{\partial \rho}
-\frac{(l-1)(l+2)}{2\rho^2}\right]j_l(k\rho). \nonumber \\
\eeqa
We extract the space-dependent part of the vector-type modes given by Eqs.~\eqref{eq:vectorpertrhoi} 
and \eqref{eq:vectorpertij} as
\begin{eqnarray}
H_{\rho i}^{(V;klm)} & = & A_V^{kl}j_l(k\rho)\mathbb{V}_i^{(lm)},\\
H_{ij}^{(V;klm)} & = & - \frac{2k_V A_V^{kl}\rho^2 \mathbb{V}_{ij}^{(lm)}}{(l-1)(l+2)}
\left( \frac{\partial\ }{\partial\rho} + \frac{2}{\rho}\right)j_l(k\rho).
\end{eqnarray}

The scalar plane wave propagating in the $z$-direction can be expanded as follows:
\begin{equation}
e^{ik z}
 =  \sum_{l=0}^\infty (2l+1)i^l j_l(k\rho)\mathrm{P}_l(\cos\theta).
\end{equation}
The space-dependent part of a circularly polarized gravitational plane wave propagating in the $z$-direction can be given as
\begin{eqnarray}
H_{xx}^{\textrm{pl}} = - H_{yy}^{\textrm{pl}} & = & \frac{1}{2}e^{ikz}, \label{g-plane1}\\
H_{xy}^{\textrm{pl}} & = & \pm \frac{i}{2}e^{ikz}. \label{g-plane2}
\end{eqnarray}
By the standard coordinate transformation of a tensor, we find
\beqa
H_{\rho\rho}^{\textrm{pl}} & = & \frac{1}{2}\sin^2\theta\,e^{\pm 2i\phi}e^{ikz} \nonumber \\
& = & \frac{1}{2}\sum_{l=0}^\infty (2l+1)i^l j_l(k\rho)\sin^2\theta \textrm{P}_l(\cos\theta)e^{\pm 2i\phi}. \nonumber \\
\eeqa

Now, by repeated use of the formula
\begin{equation}
\sqrt{1-x^2}\mathrm{P}_l^m(x) = \frac{1}{2l +1}
\left[ - \mathrm{P}_{l+1}^{m+1}(x) + \mathrm{P}_{l-1}^{m+1}(x)\right],
\end{equation}
where we let $\mathrm{P}_l^m(x) = 0$ if $|m| > l$, we obtain
\begin{eqnarray}
H_{\rho\rho}^{\textrm{pl}} & = & \frac{1}{2}\sum_{l=0}^\infty
i^l j_l(k \rho) e^{\pm 2i\phi} \nonumber \\
&& \times \left\{ \frac{1}{2l+3}\left[ \textrm{P}_{l+2}^2(\cos\theta)  -\textrm{P}_{l}^2(\cos\theta)\right]\right. 
\nonumber \\
&& \left. \ \ \ \ \ \ \ \ 
- \frac{1}{2l-1}\left[ \textrm{P}_l^2(\cos\theta) - \textrm{P}_{l-2}^2(\cos\theta)\right]\right\} \nonumber \\
& = & - \frac{1}{2}\sum_{l=2}^\infty i^l \textrm{P}_l^2(\cos\theta)e^{\pm 2i\phi} \nonumber \\
&& \times \left\{ \frac{1}{2l-1}\left[j_{l-2}(k\rho)+j_l(k\rho )\right]  \right. \nonumber \\
&& \left. \ \ \ \ \ \  \ + \frac{1}{2l+3}\left[ j_l(k \rho) + j_{l+2}(k \rho)\right]\right\}.
\end{eqnarray}
Then, by using
\beq
j_{l-1}(x) + j_{l+1}(x) & = & \frac{2l + 1}{x}j_l(x), \label{jeq}
\eeq
we find
\beqa
H_{\rho\rho}^{\textrm{pl}}
& = & - \frac{1}{2k^2\rho^2}\sum_{l=2}^\infty i^l (2l+1)j_l(k\rho)
\textrm{P}_l^2(\cos\theta)e^{\pm 2i\phi} \nonumber \\
& = &  - \frac{1}{\rho^2}\sum_{l=2}^\infty
i^l \sqrt{2\pi(2l+1)}j_l(k\rho) A_S^{kl}\mathbb{S}^{(l,\pm 2)}(\theta,\phi), \nonumber \\
\eeqa
where $\mathbb{S}^{(lm)}(\theta,\phi)$ is defined by Eq.~\eqref{Slm-def} with the constant
$A_S^{kl}$ defined by Eq.~\eqref{AklS}.  By comparing this expression with Eq.~\eqref{HSklm} we find
\beq
H_{\rho\rho}^{\textrm{pl}} = - \sum_{l=2}^\infty i^l \sqrt{2\pi (2l+1)}H_{\rho\rho}^{(S;kl,\pm 2)}.
\label{scalar-contrib}
\eeq

To find the vector-type contribution to the plane wave, we note that
\beqa
H^{\textrm{pl}}_{\rho \theta} & = & \frac{\rho}{2}\sin\theta\cos\theta e^{\pm 2i\phi}e^{ik\rho\cos\theta},\\
H^{\textrm{pl}}_{\rho \phi} & = & \pm \frac{i\rho}{2}\sin^2\theta e^{\pm 2i\phi}e^{ik\rho\cos\theta}. 
\eeqa
Hence
\beqa
\epsilon^{ij}\hat{D}_i H^{\textrm{pl}}_{\rho j}
& = & \pm k\rho^2 H_{\rho\rho}^{\textrm{pl}} \nonumber \\
&  = & \mp \sum_{l=2}^\infty i^l \sqrt{2\pi (2l+1)l(l+1)}A_V^{kl} j_l(k\rho) \mathbb{S}^{(l,\pm 2)}, \nonumber \\
\label{epsilonDH}
\eeqa
where the constant $A_V^{kl}$ is given by Eq.~\eqref{AklV}.
On the other hand 
\beq
\epsilon^{ij}\hat{D}_i H_{\rho j}^{(V;klm)} = \sqrt{l(l+1)}A_V^{kl}j_l(k\rho)\mathbb{S}^{(lm)}.
\eeq
By comparing this equation with Eq.~\eqref{epsilonDH} we conclude that
\beqa
\epsilon^{ij}\hat{D}_i H^{\textrm{pl}}_{\rho j} & = &  
\mp \sum_{l=2}^\infty i^l \sqrt{2\pi (2l+1)}\epsilon^{ij} \hat{D}_i H_{\rho j}^{(V;kl,\pm 2)}. \nonumber \\
\eeqa
From this equation and Eq.~\eqref{scalar-contrib} we find
\beq
H_{\mu\nu}^{\textrm{pl}}
= - \sum_{l=2}^\infty i^l \sqrt{2\pi(2l+1)}\left[
H_{\mu\nu}^{(S;kl,\pm 2)} \pm  H_{\mu\nu}^{(V;kl,\pm 2)}\right]. \nonumber \\
\eeq

%

\end{document}